\documentclass[intlimits,twoside,a4paper]{article}
\usepackage[cp1251]{inputenc}
\usepackage{bm}

\def\bb#1{\mbox{\boldmath ${#1}$}}

\usepackage[eqsecnum]{cmpj3}

\issue{2017}{20}{3}{33601}
\doinumber{10.5488/CMP.20.33601}

\title{New solid phase of dipolar systems}
\author[D.~Levesque]{D.~Levesque }

\address{Laboratoire de Physique Th\'eorique, CNRS,
Universit\'e de Paris-Sud, Universit\'e de Paris-Saclay, \\
B\^atiment 210, 91405 Orsay Cedex, France}

\date{Received April 12, 2017, in final form May 17, 2017}

\begin{document}

\maketitle

\begin{abstract}
The systems of molecules with a permanent dipole moment have solid phases
with various crystal symmetries. In particular, the solid phases
of the simplest of these systems, the  dipolar hard sphere model,
have been extensively studied in the literature. The article presents
Monte Carlo simulation results which, at low temperature, point to the stability
of a polarized solid phase of dipolar hard spheres with
the unusual number of eleven nearest neighbors, the so-called primitive tetragonal packing
or tetragonal close packing.

\pacs 61.50.Ah, 64.70.K-, 81.30.Dz
\end{abstract}

\section{Introduction}

This work is dedicated to Jean-Pierre Badiali and our friendly collaboration.

The solid phases of dipolar spherical molecules have
been studied both by numerical simulations \cite{Wei:92a,Wei:92b,Weis:93,Weis:94} and theoretical
 approaches \cite{Fries:85,Klapp:00}. At low temperatures, these
 phases are polarized and ferroelectric. They have crystal
 symmetries seemingly induced by the strong dipolar head-to-tail interaction
 which, in polarized phases at these temperatures, modifies the unit cells of
 the body centered cubic (bcc), face centered cubic (fcc) or hexagonal close packed (hcp)
 lattices which are expected to be the stable solid phases of particles with a spherical symmetry.
 Hence, on the basis of the density functional theory, following the density, temperature
 and dipolar moment values, the body centered tetragonal (bct), body centered orthorhombic
 (bco), face centered tetragonal or orthorhombic lattices have been estimated to be
 the stable solid phases of the Stockmayer system as well as the fundamental states
 of the dipolar soft sphere system \cite{Groh:01}. Furthermore, the ground state of 
 electrorheological fluids has been investigated; in \cite{Tao:91,Dav:92},
 supposing that the ground state lattice is one among simple cubic, fcc, bcc, hcp or bct,
 it was established  that the bct lattice has a lower energy.

 The solid phases of the dipolar hard sphere (DHS) system have been studied in references
 \cite{Gao:00,Hyn:05a,Hyn:05b} by numerical simulations. However, at the solid phase density,
 in  Monte Carlo (MC) simulations realized in the canonical ($NVT$) ensemble, the constant
 value and shape of the volume with periodic boundary conditions and the hard core pair potential
 strongly preclude in the MC sampling   the
 initial configuration, of a given crystal symmetry, to evolve
 towards the configurations of a lattice with a different symmetry and lower free energy.
 The MC simulations in the isothermal constant pressure ($NpT$) ensemble where the size and shape of
 the volume with periodic boundary conditions change, seem to make
 easier the transition between the initial
 configuration and the configurations of a thermodynamically more stable lattice.
 In practice, a very tight packing of dipolar hard spheres
 in solid polarized phases slows or thwarts such an evolution.
 Furthermore, a relative stability of phases estimated from their free energies
 is not obtained from the $NVT$ or $NpT$ simulations.

 In order to circumvent this last shortcoming of the MC sampling
 in the  $NVT$ and  $NpT$ ensembles and then to remove the uncertainty
 on the relative thermodynamic stability of solid polarized phases,
 recent works \cite{Hyn:05a,Hyn:05b,Lev:12} have estimated the free energy differences between
 solid phases of specified symmetry for the DHS system.
 These studies rely on the thermodynamic integration scheme \cite{Fren:84}
 and on the Jarzynski relation \cite{Jar:97a,Jar:97b}. In the
 two approaches, the free energy difference $\Delta F_{\text{BA}}=F_{\text B}-F_{\text A}$
 between two equilibrium states A and B of a system at density $\rho$ and
 temperature $T$ is computed.
 $\Delta F_{\text{BA}}$ is given in the thermodynamic integration scheme by:
 \begin{eqnarray}
   \Delta F_{\text{BA}} =  \int_{\lambda_{\text A}}^{\lambda_{\text B}} \rd{\lambda}
    \Big\langle \frac{\partial U_\lambda}{\partial \lambda} \Big\rangle_\lambda .
     \label{b1}
\end{eqnarray}
 Here, $\langle \ldots \rangle_\lambda$ is the canonical average
 of the partial derivative with respect to $\lambda$ of the internal energy $U_\lambda$ of the system
evolving through a set of equilibrium states from the state A to state B when the parameter $\lambda$
varies from $\lambda_{\text A}$ to $\lambda_{\text B}$. Such a computation of $\Delta F_{\text{BA}}$ supposes
that the transition from $U_{\lambda_{\text A}}$ to $U_{\lambda_{\text B}}$ can be made by using a parametrisation
with a unique parameter and that the number of degrees of freedom stays identical in the two states A and B.
For the computation of solid phase free energies in \cite{Hyn:05a,Hyn:05b}, the internal energy $U_{\lambda_{\text A}}$
is that of a solid harmonic model while  $U_{\lambda_{\text B}}$ is that of a system of particles interacting
by a pair potential, the crystal symmetries of the A, intermediate and B states being supposed identical.
However, since in the harmonic solid model (state A), the particles do not have rotational
degrees of freedom, also, in state B,  the particles do not have such degrees of freedom.
Such a constraint, if the state~B corresponds to a system of polar particles, implies that the particle
dipole moments remain fixed (cf.~\cite{Hyn:05b}).

 The Jarzynski relation writes:
  \begin{eqnarray}
 \langle \exp(-W/k_{\text B} T)\rangle =  \exp(-\Delta F_{\text{BA}}/k_{\text B} T) ,
     \label{b2}
    \end{eqnarray}
in which the average $\langle \exp(-W/k_{\text B} T ) \rangle$,  where $k_{\text B}$ is the Boltzmann constant, is  estimated from
the works $W$ effectuated along a set of paths
inducing a transition between the states A and B, paths which can involve stable and metastable states.
However, clearly, the absolute free energy of state B will be obtained only when
that of the state A is known.
In \cite{Lev:12}, $\Delta F_{\text{BA}}$ is estimated from equation~(\ref{b2}) by computing the work
$W$ needed to progressively modify the volume shape of systems
with periodic boundary conditions from an initial shape compatible with the lattice
symmetry of state A to a shape compatible with the lattice symmetry of state B.

In both computation methods of $\Delta F_{\text{BA}}$,  defects can appear in the particle arrangements
when $\lambda$ or the volume shape vary. In particular, in the $\Delta F_{\text{BA}}$ evaluation scheme
based on the Jarzynski relation, it is needed to check if the state B has
the expected symmetry. It is one of the aims of this work to discuss to what extent
this identification is unambiguous.

Section \ref{II} is devoted to presenting the details of the tests
which can be used to proceed to this identification.
In section \ref{III}, it is shown on the basis of these tests that at high density and
high dipole moment values, a DHS stable solid phase can be
of a tetragonal close packing (tcp) type.

\section{Symmetry tests} \label{II}

The MC simulations for estimating  $\langle \exp({-W/k_{\text B} T}) \rangle$
were realized for systems of parallelepipedic volume $V$
with periodic boundary conditions and  $N \sim 1000$
dipolar hard spheres interacting by the potential
\begin{equation}
\label{I1}
v(\textbf{r}_{ij},\textbf{s}_i,\textbf{s}_j)= v_{\text{hs}}(r_{ij}) +
 \frac{\mu^2}{r_{ij}^3}    \left [ \textbf{s}_i \cdot
\textbf{s}_j - \frac{3 \textbf{s}_i \cdot (\textbf{r}_{i}-\textbf{r}_{j})
 \textbf{s}_j \cdot (\textbf{r}_{i}-\textbf{r}_{j})}{r_{ij}^{2}} \right ].
\end{equation}
$v_{\text{hs}}(r)$ is a hard sphere potential of diameter $\sigma$ and
the second term on the r.h.s. of equation~(\ref{I1}) the dipolar interaction.
$\textbf{r}_i$ and $\textbf{s}_i$ are, respectively, the vector position
of the sphere $i$ and a unit vector associated with the orientation of
its dipole moment $\bb{\mu}_i=\mu \, \textbf{s}_i$, $\mu=|\bb{\mu}_i|$ and
$r_{ij}=  | \textbf{r}_{j} - \textbf{r}_{i}| $. The reduced density  $\rho^*=\rho \sigma^3$
and the reduced dipole moment $\mu^*=\mu/\sqrt{k_{\text B}T\sigma^3}$ characterize the states
of the DHS system. Furthermore, the value of $N$  and the periodic boundary conditions
must be chosen such that they are compatible with the crystal symmetry of
the initial state A and that expected for the final state B.

From the pressure tensor associated with the interparticle interactions  $v(\textbf{r}_{ij},\textbf{s}_i,\textbf{s}_j)$
which includes both the contributions of $v_{\text{hs}}(r_{ij})$ and dipolar interaction,
it is straightforward, as described in \cite{Aga:03,All:06,Mar:06},  to compute the work
needed to change the shape of a DHS system by a series of small homogeneous deformations.
These deformations generate the transition between states A and B of the same $V$, $N$
and $T$ values but with different periodic boundary conditions. Obviously, there
are many types of such paths allowing one to generate a transition between the volumes of the same size
but with different shapes. In \cite{Lev:12}, to maintain the system in a solid phase
and to minimize the formation of defects, the transition paths between the states A and B
at constant $T$ prevented the system from melting by keeping the density variation in
the limit of a few percent
with respect to the identical density of the initial states A and final states B.

The identification of the lattice symmetry for a configuration relies on the
pair distribution function~$g(r)$, the structure factor, the positions and numbers of particles
present in the neighboring shells, as well as on the types of Voronoi polyhedra
associated with the  nearest neighbors of a particle. Clearly, these properties are not independent,
but their simultaneous use allows one to reduce the ambiguities resulting
from the fluctuations in the local arrangement of the particles.

\begin{figure}[!b]
\vspace{-5mm}
\centering
\includegraphics[width=0.42\textwidth]{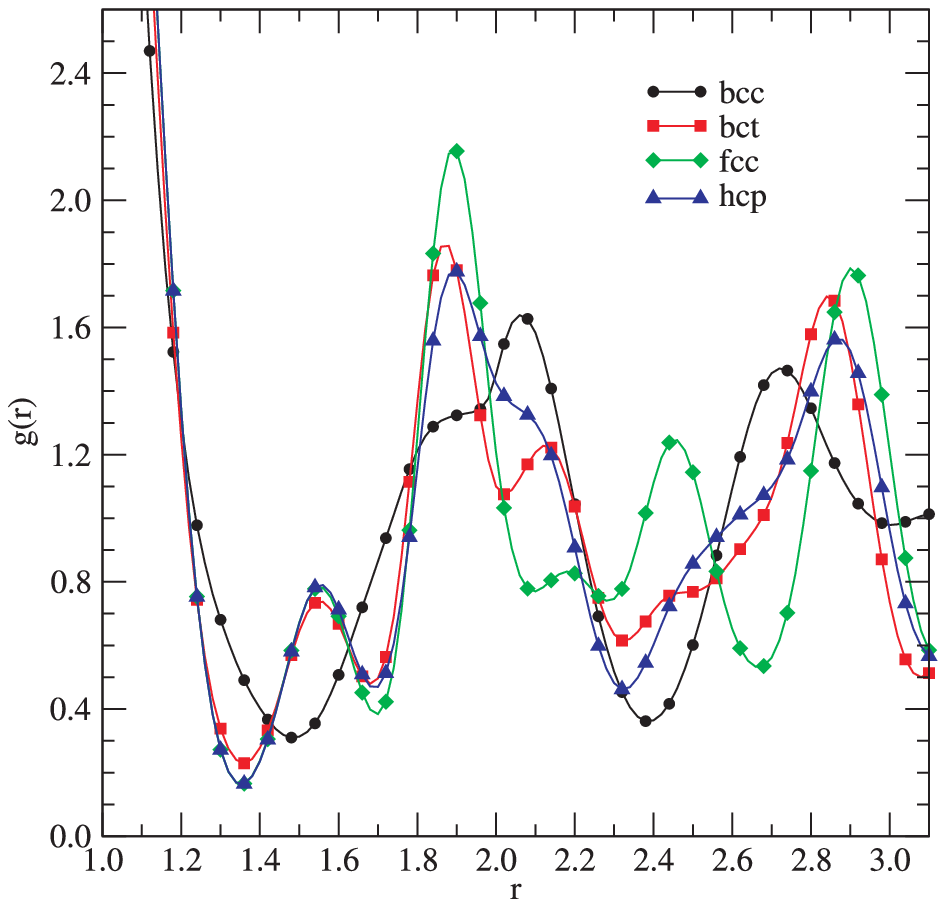}\hspace{5mm}
\includegraphics[width=0.42\textwidth]{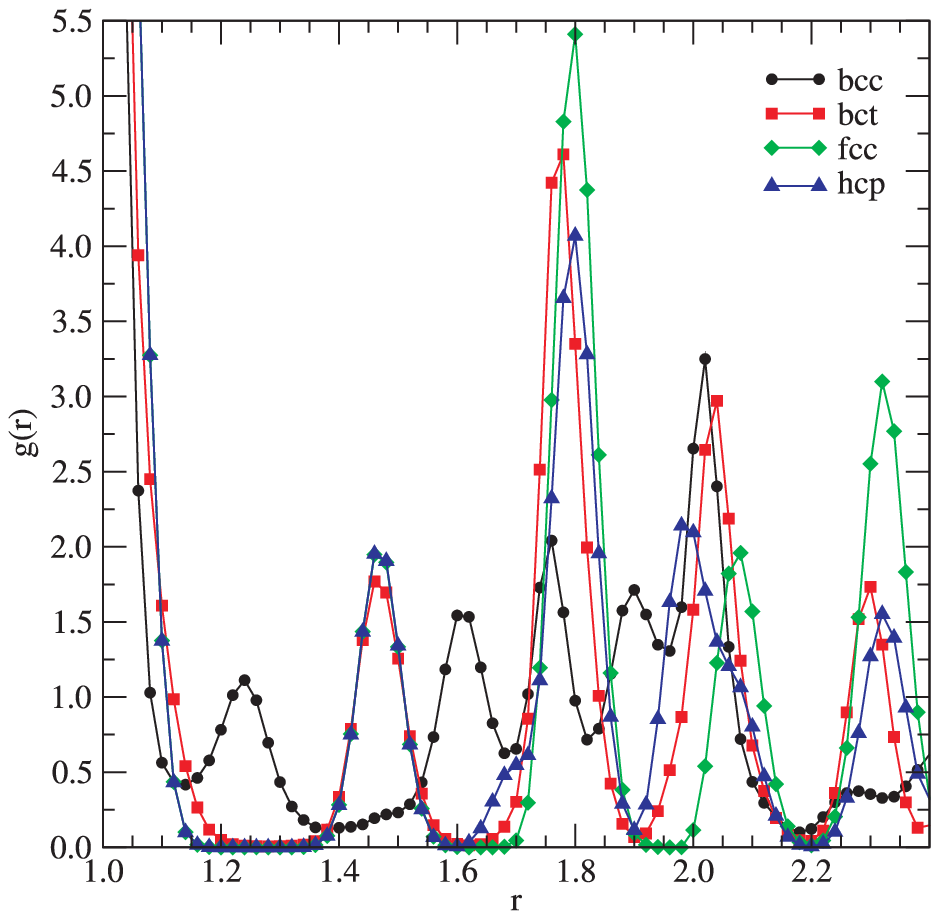}
\caption{\label{fig1} (Color online) Left-hand panel. At $\rho^*=1.06$, pair distribution functions $g(r)$ of a HS system
obtained from MC $NVT$ simulations in volumes with periodic boundary conditions
and initial configurations corresponding to bcc lattice: black dot and solid line,
 bct lattice: red square and solid line, fcc lattice: green diamond and solid line,
hcp lattice: blue up triangle and solid line.
Right-hand panel. Same at $\rho^*=1.26$. }
\end{figure}

For instance, to determine the crystal symmetry of DHS configurations at $\rho^* =1.06{-}1.26$
and  $\mu^*=1.0{-}3.0$, a simple way is to compare their two-body distribution functions
$g(r)$ with those of solid HS systems, computed by MC simulations in the canonical ensemble
for solid states of the known symmetry. Indeed, as mentioned above,
at these densities, the constant volume and hard core packing effects seem a priori to preclude
any symmetry change. Figure~\ref{fig1} shows, at these densities, such results for
$\mu^*=0.0$, i.e., a HS system, obtained after $ 40 \cdot10^6$ MC trial
moves in the $NVT$
ensemble, when, in the initial configurations, the HS are located on a bcc, bct, fcc or hcp
lattice. The functions $g(r)$ of equilibrium bcc  configurations at both densities clearly differ
from those computed for the other crystal symmetries, while the
positions of the two first peaks in $g(r)$
at $r<1.6$ for the bct, fcc and hcp  configurations are almost identical, the third peaks
at $r= 1.8{-}1.9$ differing only by their amplitudes. It is only for $r > 1.9$
that, in particular, the $g(r)$ functions of the fcc solid are clearly distinct from those
of the bct and hcp solids, whereas the bct and hcp $g(r)$ functions stay  more similar.

Additional information to interpret these $g(r)$ data can be obtained by computing the number
of neighbors $n_k$ involved in the $g(r)$'s $k$-th peak and, also, the average locations
of the neighbors around one HS in these equilibrium solid configurations.
$n_k$ is defined by
\begin{equation}
\label{I6}
 n_k  =  4 \piup \rho \int_{r^{\text{inf}}_k}^{r^{\text{sup}}_k} g(r) r^2 \rd r,
\end{equation}
where ${r^{\text{inf}}_k}$ and ${r^{\text{sup}}_k}$ are the positions of the  $g(r)$ local minima
surrounding its $k$-th peak.
For a configuration of $N$  HS, the average position
$\bar r_i$ of the $i$-th neighbor can be defined as:
\begin{equation}
\label{I2}
\bar r_i = \frac {\sum_{j=1,N} |\textbf{r}_j - \textbf{r}^j_i| }{N}\,,
\end{equation}
\looseness=-1 where $\textbf{r}_j$ is the position of the HS particle $j$, while $\textbf{r}^j_i$ is that
of its $i$-th neighbor sorted into the increasing values of
 $|\textbf{r}_j - \textbf{r}^j_i|$, so $i=1$ is the nearest neighbor,
$i=2$  the second nearest neighbor, $\ldots.$ These average positions
indicate whether the numbers of particles in the $g(r)$ peaks are in agreement with
those expected in the neighboring shells for the symmetry of the considered solid phase.
Indeed, due to the density fluctuations, the  neighbor average positions
in a shell present a dispersion reflecting the $g(r)$ peak width and, in the $\bar r_i$
graphs, the neighboring shells are indicated by a gap or a
variation of the curves slope.

\begin{figure}[!b]
\centering
\includegraphics[width=0.94\textwidth]{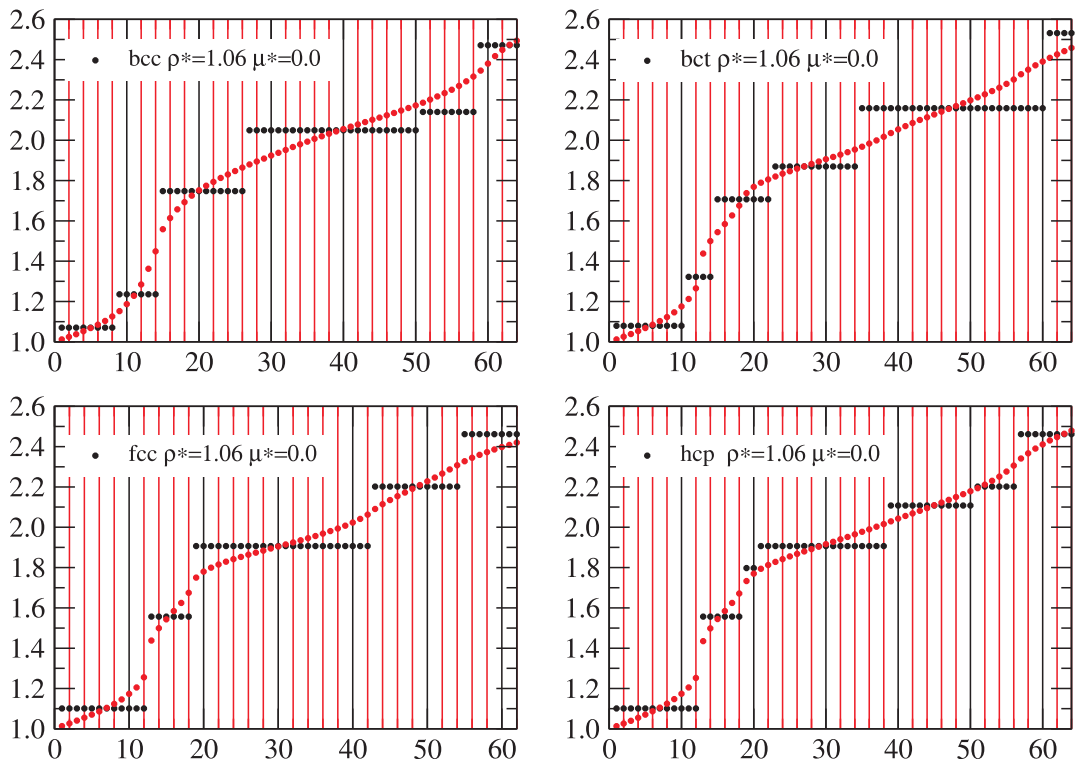}
\caption{\label{fig2} (Color online) For a HS system  at $\rho^*=1.06$, comparison of the $\bar r_i$ values
for bcc, bct, fcc and hcp configurations: black dots perfect lattice, red dots: equilibrium configuration.}
\end{figure}
\begin{figure}[!t]
\centering
\includegraphics[width=0.94\textwidth]{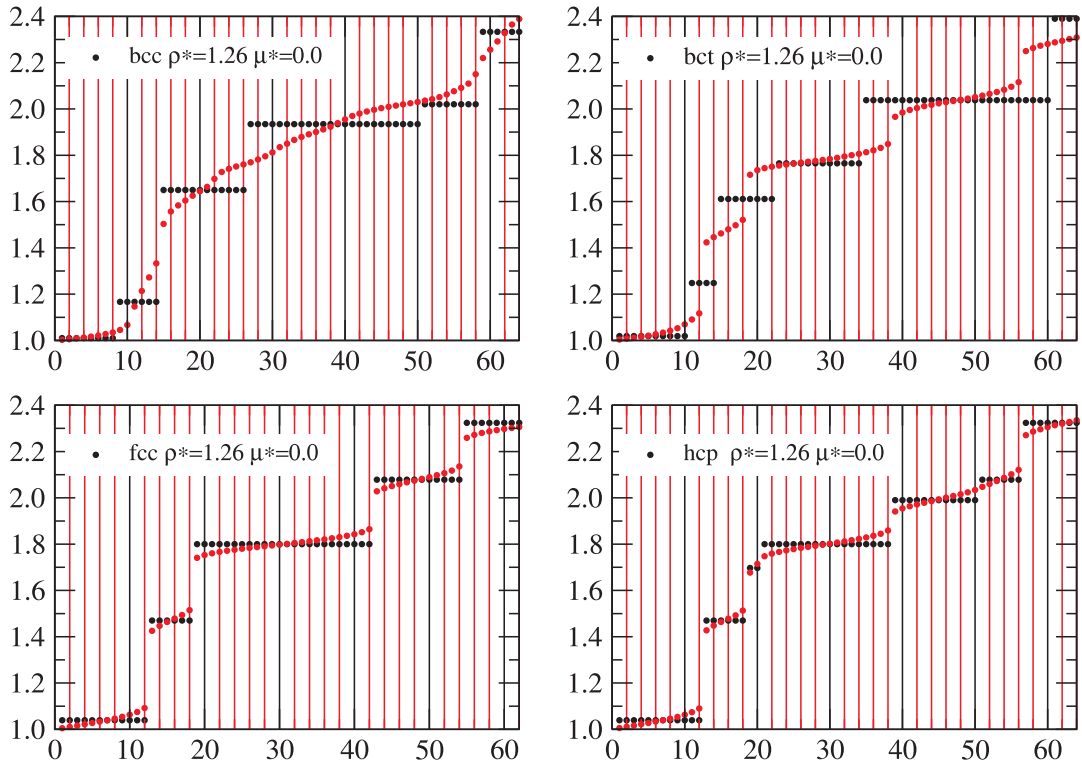}
\caption{\label{fig3} (Color online) For a HS system  at $\rho^*=1.26$, comparison of the $\bar r_i$ values
for bcc, bct, fcc and hcp configurations: black dots perfect lattice, red dots: equilibrium configuration. }
\end{figure}

Results of $\bar r_i$ computations are presented in figures~\ref{fig2} and \ref{fig3} for the equilibrium HS
configurations at $\rho^*=1.06$ and $1.26$ obtained from HS initial configurations located
on a bcc, bct, fcc or hcp lattice. At $\rho^*=1.06$, figure~\ref{fig2} clearly shows the important dispersion
of the first, second and third neighbor shells for the noncompact initial bcc and bct configurations.
From the $\bar r_i$ plot for the bcc equilibrium configurations, up to $2.5$, it appears
that the  two first shells overlap and the $4$ to $6$ th shells are largely spread. Furthermore,
the $n_k$ values indicate that the $g(r)$'s first peak involves 14 neighbors and the second
peak involves $46$ neighbors. Similarly, in the bct configuration, the two first shells overlap,
the $g(r)$ first peak involving 12 neighbors and the second, third and fourth peak
6, 20 and 18 neighbors, respectively. In the compact fcc and hcp
configurations, the two first peaks in $g(r)$
correspond to 12 and 6 neighbors in agreement with the values expected for a perfect crystal.
This agreement holds for the  third,
fourth and fifth shells in the fcc configuration in spite of the shell broadening. In the hcp configuration, the third to sixth shells overlap
to form the $g(r)$ third peak. For this solid low density, figures~\ref{fig1} and \ref{fig2} give
clear evidence
that in the bct configuration, the local structure of the three first neighboring shells
rearranges towards that found for the hcp configuration.

At $\rho^*=1.26$ compared to $\rho^*=1.06$, the shell spreading is reduced. In the fcc $g(r)$ plot,
 the peak positions and widths correspond to the positions and expected numbers of neighbors
 of a  perfect fcc solid up to $r > 3.0$. For the hcp configuration, the $g(r)$ third and fourth peaks
 result from the overlaps of the third and fourth shells, and of the fifth and
 sixth shells, respectively.
 Clearly, the HS local arrangement in the equilibrium bct configuration is very similar, almost identical
 to that obtained in the equilibrium hcp configurations. In the equilibrium bcc configurations, the $g(r)$
 first and second peaks involve together $\sim 14$ neighbors, about $10{-}11$ in the first and $3{-}4$ in the second
 peak, these neighbor numbers and the  position of the second peak being similar to those expected
 in a bct crystal.

The Voronoi tessellation of the HS considered solid configurations give little additional information
on their lattice symmetry. In the fcc and hcp configurations, at both densities, the face, edge and vertice
numbers of the Voronoi polyhedra significantly differ for most HS from the values characterising
the polyhedra in the fcc and hcp perfect lattices. A similar result is obtained for the bct configurations.
It is only for the bcc configuration that the 14 HS neighbors, present in the $g(r)$  first peak at  $\rho^*=1.06$
and the first and second  peaks at  $\rho^*=1.26$, delimit, for more than 50\% of HS, a Voronoi polyhedra with
face, edge and vertice numbers in agreement with the polyhedra of a bcc perfect lattice.

The analysis of HS solid configurations with regard to the identification of
their crystal symmetry, shows that in solid noncompact configurations, the HS local arrangement,
compared to that of perfect lattices, can be strongly modified by a large overlap of the neighboring shells.
These modifications can induce an evolution of the local arrangement towards that of more stable configurations,
as seems to happen for the bct configurations, where the local arrangement becomes of hcp type
at the two considered densities. Noticeably, in spite of the nearest neighbor shell spreading, in the tessellation
of the bcc configurations, most Voronoi polyhedra have 6 faces with 4 edges and 8 faces with 6 edges
disposed as in the polyhedra of the bcc and bct perfect lattices.

In conclusion, the association of $g(r)$ data with computed values of $n_k$ and $\bar r_i$  and, possibly,
the Voronoi tessellation allows one to obtain, at the equilibrium in a solid phase, a definite characterization
 of the HS local arrangement, but, of the crystal symmetry only for HS compact configurations.

\section{Symmetry of DHS polarized solids} \label{III}

The previous analysis made on the HS solid configurations is used now to study the symmetries
of the DHS configurations at $\rho^*=1.26$ with $\mu^* \neq 0$, the aim being to achieve a better
characterization of these symmetries than in \cite{Lev:12}. In the latter work, as mentioned above,
the symmetries were supposed to be those associated with the periodic boundary conditions of the volume $V$ enclosing
the DHS system in the final states B of the transition paths used to compute $\Delta F_{\text{BA}}$
by equation~(\ref{b2}). In these computations, $N$, the numbers of DHS in $V$ and periodic boundary conditions
were chosen in such a way that the B states can be bct, bco, fcc or hcp configurations without defects.
By comparing the $\Delta F_{\text{BA}}$ values obtained with respect to an identical initial A state, namely a
bcc configuration at equilibrium, it was
possible to estimate the relative stability of the  B states of different symmetries.

\begin{figure}[!b]
\centering
\includegraphics[width=0.48\textwidth]{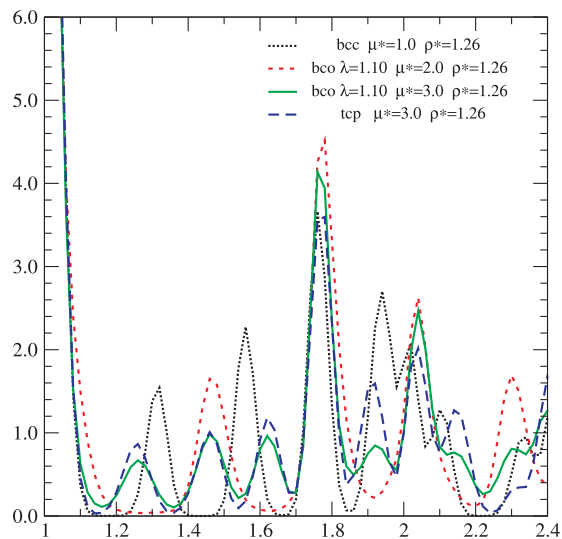}
\caption{\label{fig4} (Color online) At $\rho^*=1.26$ and $\mu^*=1.0$, $2.0$ and $3.0$, pair distribution
functions $g(r)$ of DHS configurations with the lowest value of $\Delta F_{\text{BA}}$.
 $\mu^*=1.0$ --- black line, $\mu^*=2.0$ --- red line, $\mu^*=3.0$ --- green line.
Blue line: $g(r)$ of tcp equilibrium configuration obtained by MC $NVT$ simulation at $\mu^*=3.0$. }
\end{figure}
\begin{figure}[!t]
\centering
\includegraphics[width=0.5\textwidth]{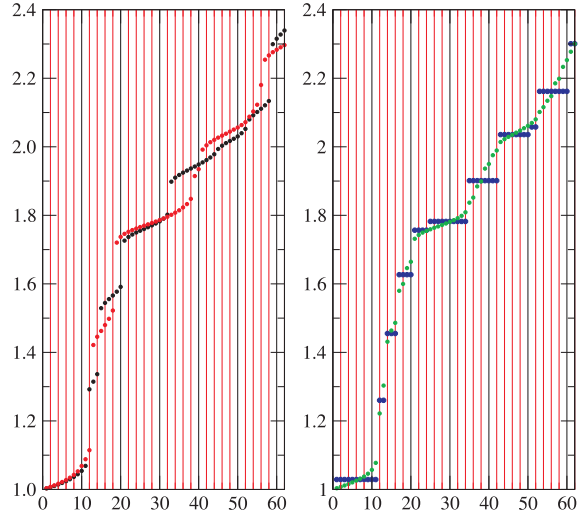}
\caption{\label{fig5} (Color online) For a DHS system  at $\rho^*=1.26$ and $\mu^*=1.0$, $2.0$ and $3.0$, comparison of the $\bar r_i$ values
for the configurations with minimal $\Delta F_{\text{BA}}$ values. Left-hand panel, black dots: bcc configuration (minimal $\Delta F_{\text{BA}}$ at $\mu^*=1.0$),
red dots: configuration (state B) obtained at the end of the transition path bcc-bco $L_y/L_x=1.10$ (minimal $\Delta F_{\text{BA}}$ at $\mu^*=2.0$).
 Right-hand panel, green dots:  configuration (state B) obtained at the end of the transition path bcc-bco $L_y/L_x=1.10$ (minimal $\Delta F_{\text{BA}}$ at $\mu^*=3.0$).
 Blue dots: $\bar r_i$ values of the tcp perfect lattice.}
\end{figure}

\looseness=-1 Hence, it was obtained that the more stable configuration at $\rho^*=1.26$ at $\mu^* = 1.0$ was bcc
and at $\mu^* =  2.0$ and $3.0$ bco with a ratio of the volume edges $L_y/ L_x$
along the $y$ and $x$ directions $\sim 1.10$. The $g(r)$ functions of such configurations
are given in figure~\ref{fig4}. In figure~\ref{fig5}, similarly to figures~\ref{fig2} and \ref{fig3}, there are given the numbers of neighbors
present in the peaks of these functions. At $\mu^*=1.0$, as in the HS system at $\rho^*=1.26$,
the first shells are modified. The first peak involves 11 neighbors and the second, third and fourth
peaks 3, 6 and 12 neighbors. This important modification of the local arrangement with respect to that of
the bcc crystal remains, as for the HS configuration, compatible with a tessellation made of Voronoi
polyhedra with 14 faces (8 with 6 edges and 6 with 4 edges) typical of the tessellation of bcc and bct
crystals. These polyhedra around a specific DHS in the configuration are strongly distorted, but when
the $N$ polyhedra are averaged, the average polyhedra are close to that of a bcc perfect crystal.
On the basis of this latter result, it seems still reasonable to label 
these unpolarized DHS configurations as bcc.

For $\mu^*=2.0$, $L_y/L_x=1.10$ and $\rho^*=1.26$,
the bco local arrangement is formed, in the perfect crystal, by shells of 2, 8, 2, 2, 4, 4, 8, 4, $\ldots$
neighbors, which are located at 1.019, 1.021, 1.190, 1.309, 1.567, 1.659, 1.766, 1.769, $\ldots.$
It is characterized, possibly very close, by shells of only a few neighbors which are expected
largely to overlap due to the local fluctuation densities. Figure~\ref{fig4}  shows how the
noncompact bco structure
is modified by the DHS dipolar interaction. The values of the number of neighbors in the first $g(r)$
peaks are equal to 12, 6, 20, 18, 14, 16, $\ldots$ values close to those obtained up to the fifth peak
for the HS $g(r)$ hcp function at $\rho^*=1.26$: 12, 6, 20, 18, 12, 20, $\ldots.$ A similar agreement
is found for the peak positions. These data show that in the DHS bco configuration at $\mu^*=2.0$ and
 $\rho^*=1.26$, similar to the HS bct configurations at this density (cf. figure~\ref{fig3}), the DHS local arrangement
evolves towards an arrangement typical of an hcp configuration.

Clearly, figure~\ref{fig4} and figure~\ref{fig5} establish how the characteristics of the equilibrium DHS configurations at $\mu^*=3.0$
and $\rho^*=1.26$ differ from those of bcc, fcc or hcp configurations. The $g(r)$ function has 8 well defined peaks
between $r=1.0$ and $r=2.2$ corresponding to the $n_k$ values of 11, 2, 3, 4, 14, 8, 10, 8~neighbors. The maximum positions of
these peaks are, respectively, 1.0, 1.26, 1.46, 1.62, 1.78, 1.92, 2.04 and 2.16. These $n_k$ values and positions
are in agreement with those expected at low temperature for a configuration,
referred to as  primitive tetragonal close packing (tcp) in the literature \cite{Baur:81,Wes:82,Dav:83}. In this type of configuration, the nine first shells correspond to
 11, 2, 3, 4, 4, 10, 8, 8, 2, 8 neighbors,
and are located, for the considered density, at 1.029, 1.260, 1.455, 1.627, 1.757, 1782, 1.901, 2.036, 2.058, and 2.161.
The fifth and sixth shells and eighth and ninth shells, considering their respective very close positions, are expected,
at finite temperature, to overlap and, hence, leading to the 14 and 10 neighbors in the $g(r)$ sixth and eighth peaks.
The $g(r)$ computed by MC  simulation  for DHS system in tcp configurations and plotted in figure~\ref{fig5}, unambiguously confirms 
a full agreement between the $g(r)$ functions of a tcp solid and those of
the states B with bco periodic boundary condition obtained at the end of the transition paths.

In \cite {Lev:12}, it was concluded from $\Delta F_{\text{BA}}$ computations that these configurations with local tcp symmetry
were the most stable among the equilibrium configurations with bct, bco, fcc and hcp periodic boundary conditions
obtained from transition paths starting from bcc configurations. In addition, in this reference, this result was supported
by the fact that in $NpT$ simulations, at $\mu^*=3.0$ and $\rho^*=1.26$, the bct initial configurations also evolved
to configurations with $g(r)$ functions similar to those of a tcp solid (cf. figure~\ref{fig4} of \cite{Lev:12}).

From this analysis of the local arrangements made on the configurations
estimated to be the most stable at $\mu^*=3.0$
and $\rho^*=1.26$, it is possible to conclude that in this domain of density and $\mu^*$ value, the DHS solid
phase has a tcp symmetry.

\section{Conclusion}

In section~\ref{II}, it was shown that the identification of the lattice symmetry of noncompact HS solid phases
made only on the basis of the $g(r)$ functions is not obvious. The local arrangement in such phases around
a given HS, presents large overlaps of the nearest neighbor shells, so that the $n_k$ values of the $g(r)$ peaks
for $r<2.0$ do not correspond to those expected in bcc or bct solid phases, but rather those of fcc or hcp phases.
However, to determine if these HS noncompact configurations eventually would evolve towards compact configurations
without defects of fcc or hcp type, prohibitively long simulations seem required. A similar uncertainty is encountered
to characterise the lattice symmetry of DHS bct or bco configurations, as
shown above for the DHS systems at $\mu^*=1.0$ and $2.0$ and $\rho^*=1.26$. Thus, as already mentioned, in \cite{Lev:12}, these
configurations with the lower values of $\Delta F_{\text{BA}}$  were labelled according to the periodic boundary conditions
of the states terminating the transition paths. Then, it is remarkable that, for $\mu^*=3.0$ and $\rho^*=1.26$,
on the basis of the excellent agreement between the $g(r)$, $n_k$ and $\bar r_i$ values with those of tcp solid phase,
it can be concluded almost unambiguously that this phase is a stable phase of the DHS system.

\ukrainianpart

\title{Нова тверда фаза дипольних систем}
\author{Д. Левек }

\address{Лабораторія теоретичної фізики, Національний центр наукових досліджень Франції,
Університет Парі-Сюд, Університет Парі-Саклє, Орсе, Франція}

\makeukrtitle

\begin{abstract}
Системи молекул з постійним дипольним моментом мають тверді фази з   кристалами різної симетрії.
Зокрема, тверді фази найпростішої з таких систем, моделі дипольних твердих сфер,
широко вивчені у літературі. В статті представлені результати моделювання методом Монте-Карло, які при низькій температурі вказують на стабільність поляризованої твердої фази дипольних твердих сфер з незвичайним числом одинадцяти найближчих сусідів, так званої примітивної тетрагональної упаковки або тетрагональної закритої упаковки.
\keywords комп'ютерне моделювання, тверді фази, дипольна тверда сфера
\end{abstract}

\end{document}